# Deep-subwavelength magnetic-coupling-dominant stereometamaterials

Zhen Gao[†1], Fei Gao[†1], Youming Zhang[1], Baile Zhang[*1,2]

[1]*Division of Physics and Applied Physics, School of Physical and Mathematical Sciences, Nanyang Technological University, Singapore 637371, Singapore.*

[2]*Centre for Disruptive Photonic Technologies, Nanyang Technological University, Singapore 637371, Singapore.*

*†These two authors contributed equally to the work.*

*\*Corresponding Author: E-mail: blzhang@ntu.edu.sg (B. Zhang)*

## Abstract

Magnetic coupling is generally much weaker than electric Coulomb interaction. This also applies to the well-known magnetic "meta-atoms," or split-ring resonators (SRRs) as originally proposed by J. Pendry *et al*, in which the associated electric dipole moments usually dictate their interaction. As a result, stereometamaterials, a stack of identical SRRs, were found with electric coupling so strong that the dispersion from merely magnetic coupling was overturned. Recently, Pendry *et al* proposed a new concept of magnetic localized surface plasmons, supported on metallic spiral structures (MSSs) at a deep-subwavlength scale. Here, we experimentally demonstrate that a stack of these new magnetic "meta-atoms" can have dominant magnetic coupling in both of its two configurations. This allows magnetic-coupling-dominant energy transport along a one-dimensional stack of MSSs, as demonstrated with near-field transmission measurement. Our work not only applies a new type of magnetic "meta-atoms" into metamaterial construction, but also provides possibilities of magnetic metamaterial design in which the electric interaction no longer takes precedence.

The field of metamaterials was greatly stimulated by the invention of split-ring resonators (SRRs) [1], as originally proposed by J. Pendry *et al*, which can generate strong magnetic moments at a deep-subwavelength scale. However, it is well-known that SRRs also carry substantial electric moments, which often dictate the interaction between adjacent SRRs. This in turn has complicated the design for magnetic metamaterials [2-4]. For example, a stack of SRRs, termed as stereometamaterials [2] as analogous to stereoisomers in chemistry, has a crucial interplay between the magnetic coupling and electric coupling, the stronger strength of the latter leading to dispersion crossing when the spatial arrangement of SRRs is tuned. Furthermore, it has been shown that a chain of subwavelength plasmonic particles can transport electromagnetic (EM) based on electric coupling [5-11] or direct electric current exchange [12-14]. However, this energy transport phenomenon has never been demonstrated with dominant magnetic coupling.

Recently, Pendry *et al*. proposed a new concept of magnetic localized surface plamsons, magnetic dipole modes supported on metallic spiral structures (MSSs) at a deep subwavlength scale [15]. Although these new magnetic "meta-atoms" have been studied individually [15-16], a stack of them, or MSS stereometamaterials, have not been studied yet. In this paper, we first theoretically and experimentally study a dimer of stacked MSSs. We demonstrate that its resonance properties can be modified by altering the stacking symmetry between two identical MSS resonators, similar to previous SRR stereometamaterials. Analysis based on Langrangian formalism [12] shows that the magnetic interaction, rather than the electric interaction, is the dominant coupling mechanism. Second, we also demonstrate that a one-dimensional vertical chain of MSSs can efficiently transport EM energy in a deep subwavelength scale with preserved magnetic mode profile.

We first illustrate the two configurations of a MSS-dimer in Fig. 1. Each MSS in the dimer consists of four spiral arms with width $w = 0.5$ mm, which are axially wrapped by 1.5 turns with outer radius $R = 12.5$ mm. The spacing of neighboring arms at the outer radius is $d = 1.5$ mm. Although the two MSSs in the dimer are identical, configurations of symmetric (Fig.1a) and anti-symmetric (Fig.1b) stacking can be arranged. The inter-MSS distance is set as $h = 6$ mm.

To study the EM response of these MSS-dimers, numerical simulations are performed based on a commercial finite-integration time domain software CST Microwave Studio. In the numerical calculations, metal is modeled as a perfect electric conductor (PEC). As indicated in Fig. 1(a), a discrete port (dipole current) excites the resonance in the dimer, and a probe is employed to detect the local $E_z$ field at the position 1 mm above the center of the upper MSS. Simulated near-field response spectra of the two different stacking configurations compared with the case of only one MSS are shown in Fig. 2. Each stacking configuration has two observable resonances, $\omega_s^-$ and $\omega_s^+$ for symmetric stacking ($\omega_s^- = 1.88$ GHz and $\omega_s^+ = 2.18$ GHz in Fig. 2) and $\omega_a^-$ and $\omega_a^+$ for anti-symmetric stacking ($\omega_a^- = 1.67$ GHz and $\omega_a^+ = 2.46$ GHz in Fig. 2), which are split from the resonance $\omega_0$ of the single MSS ($\omega_0 = 1.99$ GHz in Fig. 2). As studied previously, this resonance $\omega_0$ of a single MSS corresponds to a magnetic dipole mode of localized surface plasmons [15-16], but how adjacent magnetic MSSs interact through magnetic coupling is still unknown.

To understand these resonance properties, we further simulate in Fig. 3 the corresponding distributions of currents, magnetic fields, and electric fields for the resonances revealed above. It can be seen that on each MSS, electric currents form a major current loop, generating a strong magnetic dipole moment [16]. For the symmetrically stacked MSS-dimer, the magnetic dipole moments ($\mu$) of the two MSSs are aligned parallel at the lower frequency $\omega_s^-$ [Fig. 3(a2)], and anti-parallel at the higher frequency $\omega_s^+$ [Fig. 3(b2)]. This is consistent with the magnetic field distributions in Fig. 3(a3) and 3(b3). In Fig. 3(a3), the magnetic fields go through these two MSSs continuously, which means the north pole of the upper MSS and the south pole of the lower MSS face each other, causing an attractive force between them. This attractive force lowers the resonance frequency from $\omega_0$ of a single MSS to $\omega_s^-$ when two MSSs are combined. On the contrary, at the higher resonance frequency $\omega_s^+$, as shown in Fig. 3(b3), the magnetic fields from these two MSSs repel in the middle region, because two like magnetic poles of MSSs face each other. This repelling force raises the resonance frequency from $\omega_0$ of a single MSS to $\omega_s^+$ when two MSSs are combined.

This resonance frequency splitting mechanism based on magnetic coupling also applies to the anti-symmetrically stacked MSS-dimer. As shown in Fig. 3(c1) and Fig. 3(d1), each MSS in this anti-symmetric stacking configuration still carries a major current loop with a strong magnetic dipole moment. At the lower resonance frequency $\omega_a^-$, the magnetic dipole moments of the upper and lower MSSs are aligned parallel [Fig. 3(c2)]. At the higher resonance frequency $\omega_a^+$, they are aligned anti-parallel [Fig. 3(d2)]. After comparing their magnetic field distributions in Fig. 3(c3) and Fig. 3(d3), we can deduce, from the same reason analyzed in the symmetric stacking configuration, that the resonance frequency splitting, from $\omega_0$ of a single MSS to $\omega_a^-$ and $\omega_a^+$ when two MSSs are combined in the anti-symmetric stacking configuration, is also because of the magnetic coupling between these two MSSs.

Despite similar magnetic coupling, the frequency differences of lower and higher resonance frequencies in the two stacking configurations are different. We need to consider the electric coupling to account for this difference. Because of the spiral geometry of MSS, the current loop [e.g. the upper MSS in Fig. 3(a1)] with a downward magnetic moment [Fig. 3(a2)] must be associated with radial currents flowing inwards, generating radially oriented electric dipoles ($p$) and accumulating positive charges at the center. Therefore, for the symmetrically stacked MSS-dimer, when the magnetic dipoles of upper and lower MSSs are aligned parallel at the lower resonance frequency $\omega_s^-$ [Fig. 3(a2)] with an attractive magnetic force, the accumulated like charges on the upper and lower MSSs repel each other and generate a repelling force, which can be inferred from the electric field distribution in Fig. 3(a4). This repelling force partially offsets the magnetic attractive force. On the other hand, at the higher resonance frequency $\omega_s^+$, as shown in Fig. 3(b2), when the magnetic dipoles of upper and lower MSSs are aligned anti-parallel, the associated radial currents for the upper and lower MSSs are anti-parallel. The accumulated opposite charges on the upper and lower MSSs generate an attractive electric force [Fig. 3(b4)], which partially offsets the repelling magnetic force [Fig. 3(b3)].

However, in the anti-symmetrically stacked MSS-dimer, the weak electric force no longer offsets the strong magnetic force. As shown in Fig. 3(c1), because of the anti-symmetric stacking, the parallelly aligned magnetic moments at the lower resonance frequency $\omega_a^-$ are associated with anti-parallel radial currents at the upper and lower MSSs [Fig. 3(c2)]. These anti-parallel radial currents will accumulate opposite charges at the center of MSSs, and generate an attractive electric force, as can be inferred from the electric field distribution in Fig. 3(c4). Similarly, at the higher resonance frequency $\omega_a^+$, the anti-parallel magnetic moments are associated with parallel radial currents on the upper and lower MSSs [Fig. 3(d2)], which accumulate like charges at the center of MSSs and generate a repelling electric force, as can be inferred from the electric field distribution in Fig. 3(d4). Therefore, in the anti-symmetric stacking, the weak electric force always enhances the strong magnetic force, leading to a wider frequency splitting for the MSS-dimer, i.e. $\omega_a^- < \omega_s^- < \omega_s^+ < \omega_a^+$.

We then quantitatively compare the magnetic coupling strength and electric coupling strength. A single MSS can be modeled as an equivalent $LC$ circuit with resonance frequency $\omega_0 = 1/(LC)^{1/2}$, where the four spiral arms carry inductance $L$ and gaps between neighboring spiral arms carry capacitance $C$. Similar to previous Lagrangian formalism for a SRR-dimer [2], we can write down the Lagrangian for this MSS-dimer system as follows:

$$\Gamma = \frac{L}{2}(\dot{q}_1^2 - \omega_0^2 q_1^2) + \frac{L}{2}(\dot{q}_2^2 - \omega_0^2 q_2^2) + M_H \dot{q}_1 \dot{q}_2 \pm M_E \omega_0^2 q_1 q_2 \quad (1)$$

where $q_m$ ($\dot{q}_m$) represents the charge (current) accumulated on the m-th MSS, and $M_H$ and $M_E$ quantify the magnetic and electric couplings, respectively. The sign of "$\pm$" before the last item takes "-" for the symmetric stacking and "+" for the anti-symmetric stacking. Substituting Eq. (1) into the Euler-Lagrange equations:

$$\frac{d}{dt}\left(\frac{\partial \Gamma}{\partial \dot{q}_m}\right) - \frac{\partial \Gamma}{q_m} = 0, (m = 1, 2) \quad (2)$$

We can obtain the eigen-frequencies of the MSS-dimer system as:

$$\begin{cases} \omega_s^- = \omega_0 \sqrt{\dfrac{1+\kappa_{Es}}{1+\kappa_{Hs}}} \\ \omega_s^+ = \omega_0 \sqrt{\dfrac{1-\kappa_{Es}}{1-\kappa_{Hs}}} \end{cases} \quad (3)$$

for the symmetric stacking configuration and

$$\begin{cases} \omega_a^- = \omega_0 \sqrt{\dfrac{1-\kappa_{Ea}}{1+\kappa_{Ha}}} \\ \omega_a^+ = \omega_0 \sqrt{\dfrac{1+\kappa_{Ea}}{1-\kappa_{Ha}}} \end{cases} \quad (4)$$

for the anti-symmetric stacking configuration, where $\kappa_{Es}(\kappa_{Ea}) = M_{Es}(M_{Ea})/L$, $\kappa_{Hs}(\kappa_{Ha}) = M_{Hs}(M_{Ha})/L$ are the electric and magnetic coupling strengths for the symmetric (anti-symmetric) stacking configuration, respectively. Fitting the simulated resonance frequencies with the above expressions yields the corresponding coupling coefficients as $\kappa_{Hs} = 0.284$, $\kappa_{Es} = 0.134$, $\kappa_{Ha} = 0.279$, and $\kappa_{Ea} = 0.105$. Notably, for both stacking configurations, the magnetic coupling strength is always much larger than the electric coupling strength.

We then proceed to experimentally demonstrate the mode splitting of the MSS-dimers. By using standard printed circuit-board fabrication, we printed 18-μm-thick MSSs on 200-μm-thick FR4 dielectric substrate. A foam plate (ROHACELL 71 HF, relative permittivity $\varepsilon = 1.075$, thickness $h = 6$ mm) was used as a spacer between two identical MSSs. The near-field response spectra were measured with a pair of monopole antennas as the source and the probe, as indicated in Fig. 1(a), connected to a vector network analyzer (R&S ZVL-13). The measured results are shown in Fig. 4(a). It can be seen that the resonance peak of a single MSS ($\Omega = 1.82$ GHz) is split into two distinct resonance peaks at frequencies $\omega_s^0 = 1.71$ GHz and $\omega_s^\pi = 1.95$ GHz for symmetric stacking, and $\omega_a^\pi = 1.51$ GHz and $\omega_a^0 = 2.18$ GHz for anti-symmetric stacking, respectively, where subscript '$\pi$' and '0' denote the relative phase between the electric fields of two MSSs. Note that because of the 200-μm-thick FR4 substrate, all resonance peaks are red-shifted slightly compared to simulation results in Fig. 2. We further present the measured near-field distributions ($E_z$) on the top and bottom transverse planes of the MSS-dimers for the symmetric stacking configuration [Fig. 4(b1-b2)] and the anti-symmetric stacking configuration [Fig. 4(c1-c2)]. Clearly, the phase difference between top and bottom MSS in the symmetric stacking configuration is $\Delta\emptyset = 0$ at the lower resonance frequency ($\omega_s^0$) [Fig. 4(b1)] and $\Delta\emptyset = \pi$ at the higher resonance frequency ($\omega_s^\pi$) [Fig. 4(b2)]. Yet the phase difference between the top and bottom MSS in the anti-symmetric stacking configuration is $\Delta\emptyset = \pi$ at the lower frequency ($\omega_a^\pi$) [Fig. 4(c1)] and $\Delta\emptyset = 0$ at the higher frequency ($\omega_a^0$) [Fig. 4(c2)]. Note that although the electric interaction enhances the magnetic interaction in anti-symmetric MSS-dimer and offsets the magnetic interaction in symmetric MSS-dimer, the sign of the total coupling coefficients are still determined by the magnetic interaction, thus the sign of the total coupling coefficient will not reverse [13].

Using an optical analogue of the tight-binding (TB) model [17-18], we can extract the measured coupling factor $\kappa$ as $|\kappa_a| = \left|\frac{\Delta\omega_a}{\Omega}\right| = \left|\frac{\omega_a^\pi - \omega_a^0}{\Omega}\right|$ for anti-symmetric stacking configuration. After substituting $\omega_a^0 = 2.18$ GHz, $\omega_a^\pi = 1.51$ GHz, and $\Omega = 1.82$ GHz, we can obtain that $\kappa_a = -0.37$. Similarly, we can obtain the coupling factor for the symmetric stacking configuration as $\kappa_s = -0.13$. It is evident that anti-symmetric coupling configuration have a larger total coupling strength.

When the number of vertically stacked magnetic MSS is increased, a waveguide transmission band can be expected due to the vertical coupling of individual resonant modes, similar to previous coupled resonator optical waveguides (CROWs) [17]. Note that previous CROWs were predominantly based on electric coupling or electric current exchange [11-12], but the CROW we will demonstrate here is magnetic-coupling dominant. The intrinsic dispersion relation of an infinite chain of vertically stacked magnetic MSS can be obtained as: $\omega_{s(a)} = \omega_0[1 + \kappa_{s(a)} \cos(Kh)]$ [17], where $\omega_0$ is the resonance frequency of an individual MSS, $|\kappa_{s(a)}| = \left|\frac{\omega_{s(a)}^- - \omega_{s(a)}^+}{\omega_0}\right| = \left|\kappa_{Hs(a)} \mp \kappa_{Es(a)}\right|$ is the inter-MSS coupling strength, $K$ is the Bloch wavevector and $h$ is the periodicity of the vertical waveguide. We calculate dispersion curves of an infinite vertical chain of MSSs arrayed in both symmetric and anti-symmetric configurations with periodicity $h = 6$ mm, as shown in Fig. 5(a). The magnetic localized surface plasmon modes are transported in the vertical waveguide in the frequency bands centered at the resonance frequency of an individual MSS (black dashed line). The transmission band of the anti-symmetric stacking configuration (red line) is wider than that of the symmetric stacking configuration (blue line), because of the larger coupling strength between neighboring MSS in the anti-symmetric stacking configuration.

To experimentally demonstrate the vertical transport of magnetic localized surface plasmons in deep-subwavelength scales, we constructed a vertical stacked CROW which consists of nine magnetic MSSs with periodicity of $h = 6.0$ mm, as shown in the bottom right inset of Fig. 5(b). The measured results are shown in Fig. 5(b). We first measured the transmission spectrum (black triangles) without the CROW showing that EM energy cannot transport vertically in air at frequencies from 0.8 GHz to 3 GHz. Then we measured the transmission spectra of two vertical magnetic CROWs with symmetric stacking configuration (blue squares) and anti-symmetric stacking configuration (red circles), respectively. Each stacking configuration exhibits a transmission band, whose frequency range matches well with the calculated band structure [Fig. 5(a)]. The measured field pattern of $E_z$ at 1.75 GHz [upper-left inset of Fig. 5(b)] on the top surface of both symmetric and anti-symmetric stacking configurations shows a clear magnetic dipole mode [15-16], showing that the mode profile is not changed during energy transport.

In conclusion, we have experimentally demonstrated a deep-subwavelength MSS-dimer system in which the magnetic interaction dominates. Lagrangian analytical model is adopted to analyze the coupling mechanism. Moreover, two deep-subwavelength vertically-coupled magnetic resonator waveguides are experimentally realized at microwave frequencies by placing magnetic MSSs in a vertical chain with two distinct stacking configurations. This work applies a new type of magnetic "meta-atoms" into metamaterial construction, and provides possibilities of magnetic metamaterial design in which the electric interaction no longer takes precedence.

**Acknowledgements**

This work was sponsored by the NTU-NAP Start-Up Grant, Singapore Ministry of Education under Grant No. Tier 1 RG27/12 and MOE2011-T3-1-005.

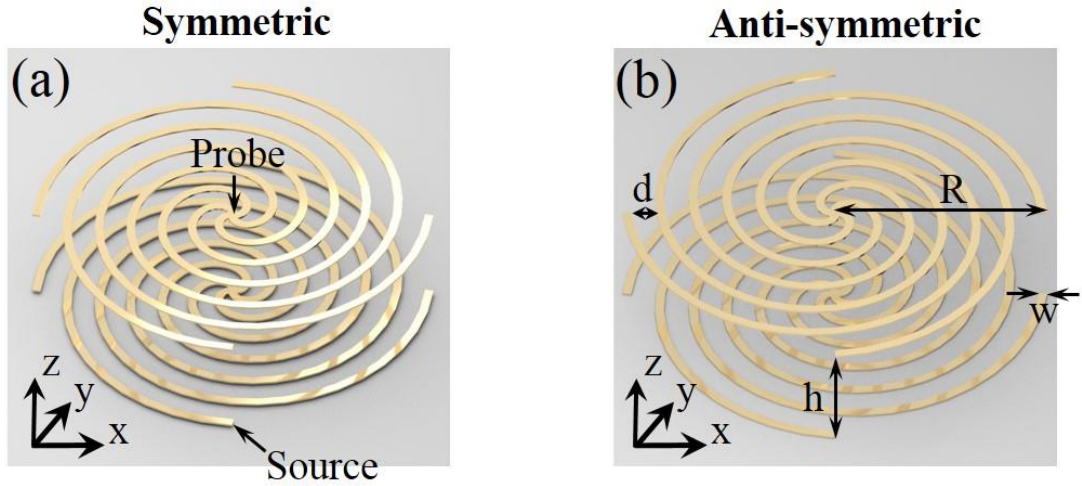

FIG. 1 (color online). Schematic of symmetric (a) and anti-symmetric (b) stacking configurations of two vertically-coupled metallic spiral structures. Two monopole antennas indicated by two black arrows are used to excite and probe the resonances. The ultrathin metallic spiral structure (thickness 0.018 mm) has outer radius $R = 12.5$ mm. The width of spiral strips is $w = 0.5$ mm. The spacing between neighboring stripes is $d = 1.5$ mm at the outer radius.

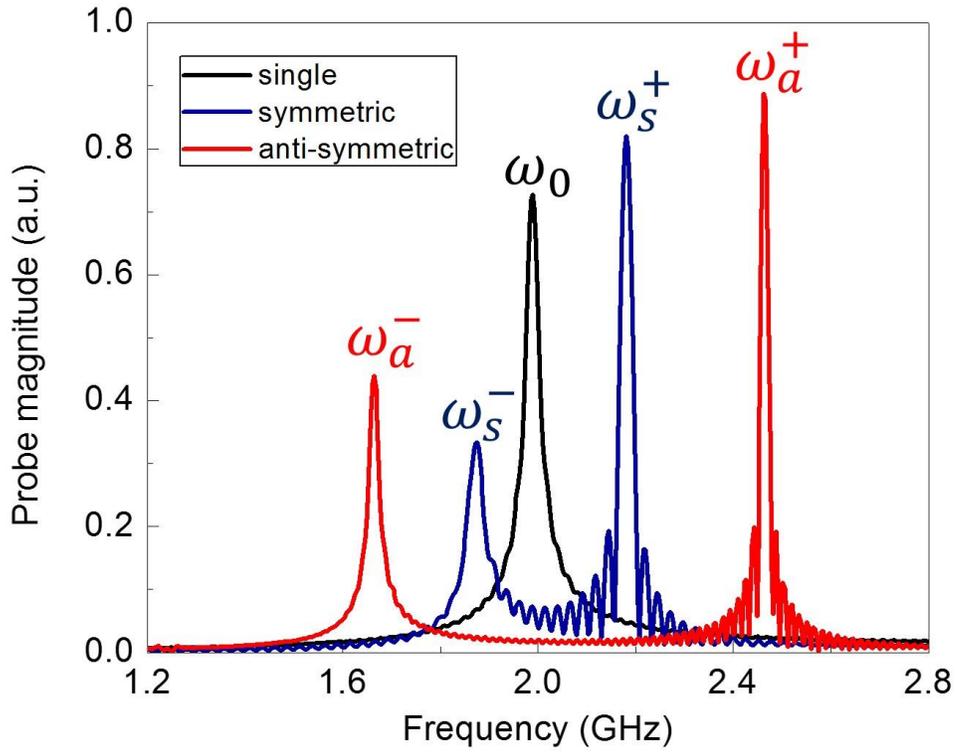

FIG. 2 (color online). Simulated near-field response spectra of two vertically-coupled metallic spiral structures with symmetric stacking configuration (blue line) and anti-symmetric stacking configuration (red line), compared with the near-field response spectrum of a single metallic spiral structure (black line).

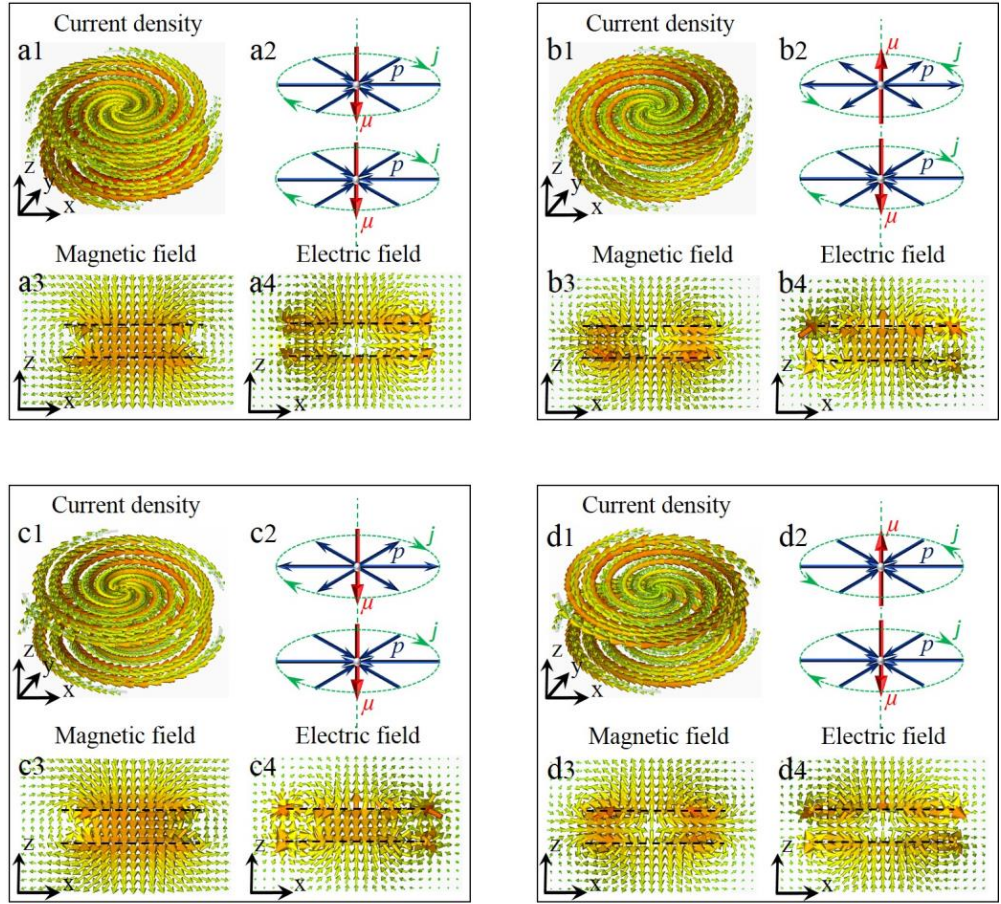

FIG. 3 (color online). Resonance modes analysis on the two vertically-coupled metallic spiral structures. (a1-a4) The current distribution (a1), the alignment of the magnetic and electric dipoles (a2), the magnetic field distribution (a3), and the electric field distribution (a4) at the lower resonance frequency ($\omega_s^-$) of symmetric stacked dimer. (b1-b4) The current distribution (b1), the alignment of the magnetic and electric dipoles (b2), the magnetic field distribution (b3), and the electric field distribution (b4) at the higher resonance frequency ($\omega_s^+$) of symmetric stacked dimer. (c1-c4) The current distribution (c1), the alignment of the magnetic and electric dipoles (c2), the magnetic field distribution (c3), and the electric field distribution (c4) at the lower resonance frequency ($\omega_a^-$) of anti-symmetric stacked dimer. (d1-d4) The current distribution (d1), the alignment of the magnetic and electric dipoles (d2), the magnetic field distribution (d3), and the electric field distribution (d4) at the higher resonance frequency ($\omega_a^+$) of anti-symmetric stacked dimer.

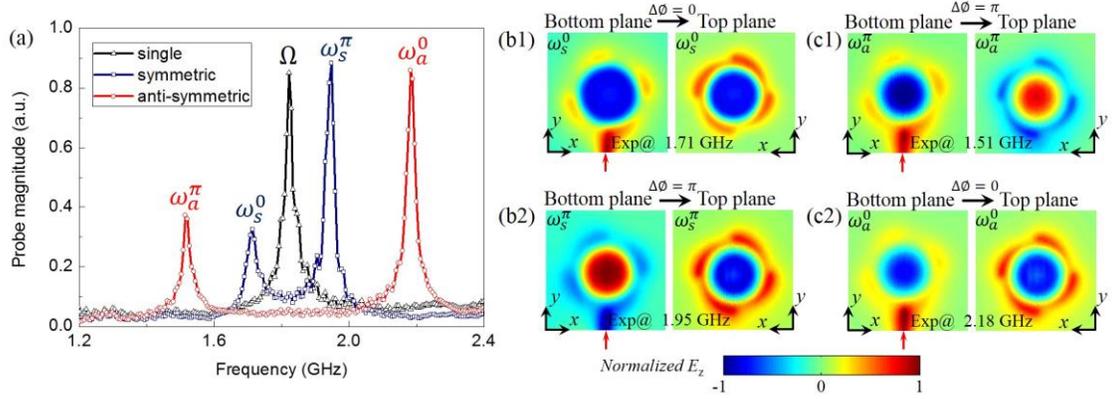

FIG. 4 (color online). (a) The measured near-field response spectra of a single metallic spiral structure (black triangles), two vertically-coupled metallic spiral structures with symmetric (blue squares) and anti-symmetric (red circles) stacking configurations. (b1-b2) The measured $E_z$ field patterns on two transverse planes 1 mm above and below the stacked dimer with symmetric stacking configuration at lower resonance frequency $\omega_s^0$ (1.71 GHz; b1) and higher frequency $\omega_s^\pi$ (1.95 GHz; b2). (c1-c2) The measured $E_z$ field patterns on two transverse planes 1 mm above and below the stacked dimer with anti-symmetric stacking configuration at lower resonance frequency $\omega_a^\pi$ (1.51 GHz; c1) and higher frequency $\omega_a^0$ (2.18 GHz; c2).

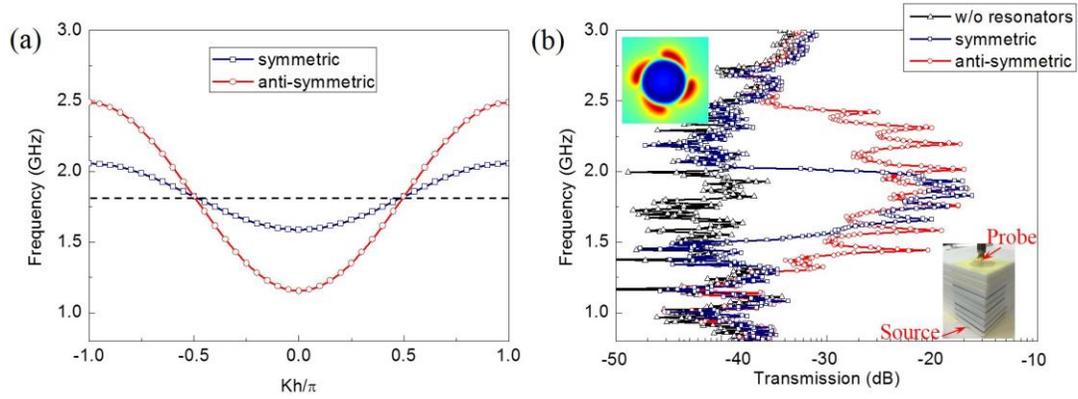

FIG. 5 (color online). (a) Dispersion diagrams of two infinite chains of metallic spiral structures with symmetric (blue line) and anti-symmetric (red line) stacking configurations. (b) Measured transmission spectra for two finite vertical magnetic coupled resonator waveguides which consist of nine MSSs separated with 6-mm thick foam plates with symmetric (blue line) and anti-symmetric stacking configurations (red line). Bottom right inset shows the experimental setup. Upper left inset shows the scanned $E_z$ field pattern at 1.75 GHz on top of the symmetric stacking waveguide.